\documentclass[aps,prd,notitlepage,showpacs,nofootinbib]{revtex4-1}

\usepackage{graphicx}

\usepackage[utf8]{inputenc} 

\usepackage{amsmath}
\usepackage{amssymb}
\usepackage{float}
\usepackage{comment}
\def \znbb {$\rm 0\nu\beta\beta$ }

\def\gsim{\raise0.3ex\hbox{$\;>$\kern-0.75em\raise-1.1ex\hbox{$\sim\;$}}}
\def\lsim{\raise0.3ex\hbox{$\;<$\kern-0.75em\raise-1.1ex\hbox{$\sim\;$}}}
\def\lfv{lepton flavour violation }

\usepackage{soul}

\usepackage[usenames,dvipsnames]{color}
\usepackage[colorinlistoftodos]{todonotes}

\definecolor{mightnightblue}{RGB}{25,25,112}

\definecolor{brown}{rgb}{0.59, 0.29, 0.0}

\newcommand {\black} {\color{black}}

\def\lfv{lepton flavour violation }

\def\vev#1{\left\langle #1\right\rangle}

\def\21{$\mathrm{SU(2)_L \otimes U(1)_Y}$}
\def\lfv{lepton flavour violation }

\def\Q{\hbox{$\cal Q$ }}

\newcommand{\AddrAHEP}{AHEP Group, Institut de F\'{i}sica Corpuscular --
  C.S.I.C./Universitat de Val\`{e}ncia, Parc Cientific de Paterna.\\
  C/Catedratico Jos\'e Beltr\'an, 2 E-46980 Paterna (Val\`{e}ncia) - SPAIN}

\newcommand{\UdeA}{Instituto de F\'\i sica, Universidad de Antioquia,
  Calle 70 No. 52-21, Apartado Aéreo 1226, Medellín, Colombia.}

\newcommand{\ICTP}{Simons Associate at The Abdus Salam International Centre for Theoretical Physics (ICTP),\\
  Strada Costiera 11, 34151, Trieste, Italy.}

\usepackage{hyperref}
 \hypersetup{
     colorlinks=true,
     linkcolor= Black,
     citecolor=mightnightblue,
     urlcolor=mightnightblue
     } 

\begin{document}
\title{\color{BrickRed}Bound-state dark matter with Majorana neutrinos}
\author{M. Reig~${}^1$}\email{mario.reig@ific.uv.es}
\author{D. Restrepo~${}^{2,3}$}\email{restrepo@udea.edu.co}
\author{J. W. F. Valle~${}^1$} \email{valle@ific.uv.es, URL: 
  http://astroparticles.es} 
\author{O. Zapata~${}^2$}\email{oalberto.zapata@udea.edu.co}

\affiliation{$^1$~\AddrAHEP}
\affiliation{$^2$~\UdeA}
\affiliation{$^3$~\ICTP}


\begin{abstract}

We propose a simple scenario in which dark matter (DM) emerges as a stable neutral hadronic thermal relics, its stability following from an exact $\operatorname{U}(1)_D$ symmetry. Neutrinos pick up radiatively induced Majorana masses from the exchange of colored DM constituents.
There is a common origin for both dark matter and neutrino mass, with a lower bound for neutrinoless double beta decay. 
Direct DM searches at nuclear recoil experiments will test the proposal, which may also lead to other phenomenological signals at future hadron collider and \lfv experiments.

  \end{abstract}

\pacs{ 13.15.+g, 14.60.Pq, 14.60.St, 95.35.+d}

\maketitle

\section{Introduction}

For a number of reasons, particle dark matter (DM) candidates are usually assumed to be electrically neutral and colorless.
This has recently been challenged, however, in Ref.~\cite{DeLuca:2018mzn}, where it was suggested that dark matter could be the lightest hadron made of two stable color octet Dirac fermions \Q with mass around $12.5~\text{TeV}$.
Bound-state dark matter may also arise from new confining hypercolor interactions beyond QCD~\cite{Higaki:2013vuv,Bai:2013xga,Detmold:2014qqa}. Such QCD-like bound state dark matter models have also been considered in the context of asymmetric dark matter scenarios~\cite{Lonsdale:2017mzg,Lonsdale:2018xwd}, and may also emerge naturally in models of comprehensive unification~\cite{Reig:2017nrz}. 
A simple theory implementation of the bound-state dark matter scenario in which a QCD bound state $\mathcal{Q}\mathcal{Q}$ is formed and its stability arises from the conservation of $B-L$ symmetry, has been proposed in \cite{Reig:2018mdk}. In such picture the Diracness of neutrinos would be responsible for dark matter stability~\cite{Chulia:2016ngi}.

In this letter we propose an alternative bound-state dark matter construction in which $B-L$ symmetry is violated and neutrinos are Majorana fermions.
In such scenario dark matter is stabilized by an extra continuous Abelian dark symmetry, $\operatorname{U}(1)_D$, under which all standard model particles are assumed to be neutral. This leads to an economical alternative version of the bound-state DM picture in which neutrinos have radiative Majorana masses. 
In addition to the heavy Dirac messenger fermion, we introduce a pair of SU(2) scalar doublets, in order to ensure that at least two neutrino masses are nonzero, as needed to account for neutrino oscillation data~\cite{deSalas:2017kay,globalfit}.

\section{Bound-state dark matter with Majorana neutrinos}
\label{sec:model}

Here we propose a simple realization of the radiative seesaw neutrino mass generation picture, in which a single colored fermion Dirac messenger \Q is introduced
along with
two colored SU(2) doublet scalars $\eta_a$ ($a=1,2$), see table~\ref{tab:bmlnur}. The heavy messenger \Q is charged under $\operatorname{U}(1)_{D}$ and can act as constituent dark matter, made stable thanks to a $\operatorname{U}(1)_{D}$ symmetry.

\begin{table}[!h]
  \centering
  \begin{tabular}{|l|c|l|}
\hline
    Particles& $\operatorname{U}(1)_{D}$ & $\left(\operatorname{SU}(3)_c ,\operatorname{SU}(2)_L \right)_Y$\\\hline
$Q_{i}=\begin{pmatrix}u_L& d_L\end{pmatrix}_i^{\operatorname{T}}$ & $0$  & $ \left(\mathbf{3},\mathbf{2}  \right)_{1/6}$  \\
$\overline{u_{Ri}}$ & $0$  & $ \left(\overline{\mathbf{3}},\mathbf{1}  \right)_{-2/3}$  \\
$\overline{d_{Ri}}$ & $0$  & $ \left(\overline{\mathbf{3}},\mathbf{1}  \right)_{1/3}$  \\ \hline
 $L_i=\begin{pmatrix}\nu_L& e_L\end{pmatrix}_i^{\operatorname{T}}$                           & $0$ & $ \left(\mathbf{1},\mathbf{2}\right)_{-1/2}$  \\
$\overline{e_{Ri}}$ & $0$  & $ \left(\mathbf{1},\mathbf{1}  \right)_{1}$  \\ 
 $\mathcal{Q}_L$    & $-1$ & $ \left(\mathbf{N}_c,\mathbf{1}\right)_0$\\
 $ \overline{\mathcal{Q}_R} $ & $1$ & $ \left(\mathbf{N}_c,\mathbf{1}\right)_0$\\\hline
 $H$                           &          0  & $ \left(\mathbf{1},\mathbf{2}\right)_{1/2}$\\ 
 $\eta_a$                        &  $(-1)^a$   & $ \left(\mathbf{N}_c,\mathbf{2}\right)_{1/2}$\\
\hline
  \end{tabular}
  \caption{Left-handed fermions and scalars. }
  \label{tab:bmlnur}
\end{table}
Note that all dark sector particles carry colour. For definiteness, we assign them to the octet $\operatorname{SU}(3)_c$ representation. This ensures the viability of the bound-state DM scenario~\cite{DeLuca:2018mzn}. Notice, however, that our neutrino mass discussion also holds if they had different $\operatorname{SU}(3)_c$ transformation properties.
The $\operatorname{U}(1)_D$ symmetry implies that \Q has only a Dirac-type mass term.
Hence the new terms in the Lagragian are the following
\begin{align}
  \label{eq:lag}
  \mathcal{L}\supset&-\left[ h_i  \overline{ \mathcal{Q}_R }\widetilde{\eta}^\dagger_1 L_i + y_i  \overline{ \mathcal{Q}_L^c }\widetilde{\eta}^\dagger_2 L_i
  + M_{\mathcal{Q}}\, \overline{\mathcal{Q}_R} \mathcal{Q}_L
  +\text{h.c}\right]
-\mathcal{V}(H,\eta_a ) ~,
\end{align}
\black
where summation is implied over repeated indices, and trace over $\mathbf{N}_c=\mathbf{8}$ is implicit. The Higgs potential contains the following terms  
\begin{align}
\mathcal{V}(H,\eta_a)=&\,V(H)+V(\eta_a)+\left[ \lambda_{3\eta H}^{ab}\left( H^{\dagger}H \right)\operatorname{Tr}\left( \eta_{a}^\dagger  \eta_{b} \right)
                       +\lambda_{4\eta H}^{ab}\operatorname{Tr}\left[\left( \eta_{a}^\dagger  H\right)\left(  H^\dagger\eta_{b} \right) \right] + \text{h.c.}\right]\delta_{ab} \nonumber\\
  +&\, \operatorname{Tr} \left[ \lambda_{\eta_1\eta_2} \left( \eta_1^{\dagger}\eta_2 \right)\left( \eta_2^{\dagger}\eta_1 \right) \right]
     +\left\{ \operatorname{Tr} \left[ \lambda_{\eta_1\eta_2 H} \left( H^{\dagger}\eta_1 \right) \left( H^{\dagger}\eta_2 \right) \right] +\text{h.c} \right\}
\end{align}
\black
Note that terms like $\left( \eta^\dagger_a H\right)^2$ are also forbidden and, as a consequence, the real and imaginary parts of the scalars with nonzero dark charges are degenerate. Moreover, the CP-even and CP-odd scalars do not mix, thanks to our CP conservation assumption. 

\begin{figure}[!h]
\centering
\includegraphics[scale=0.5]{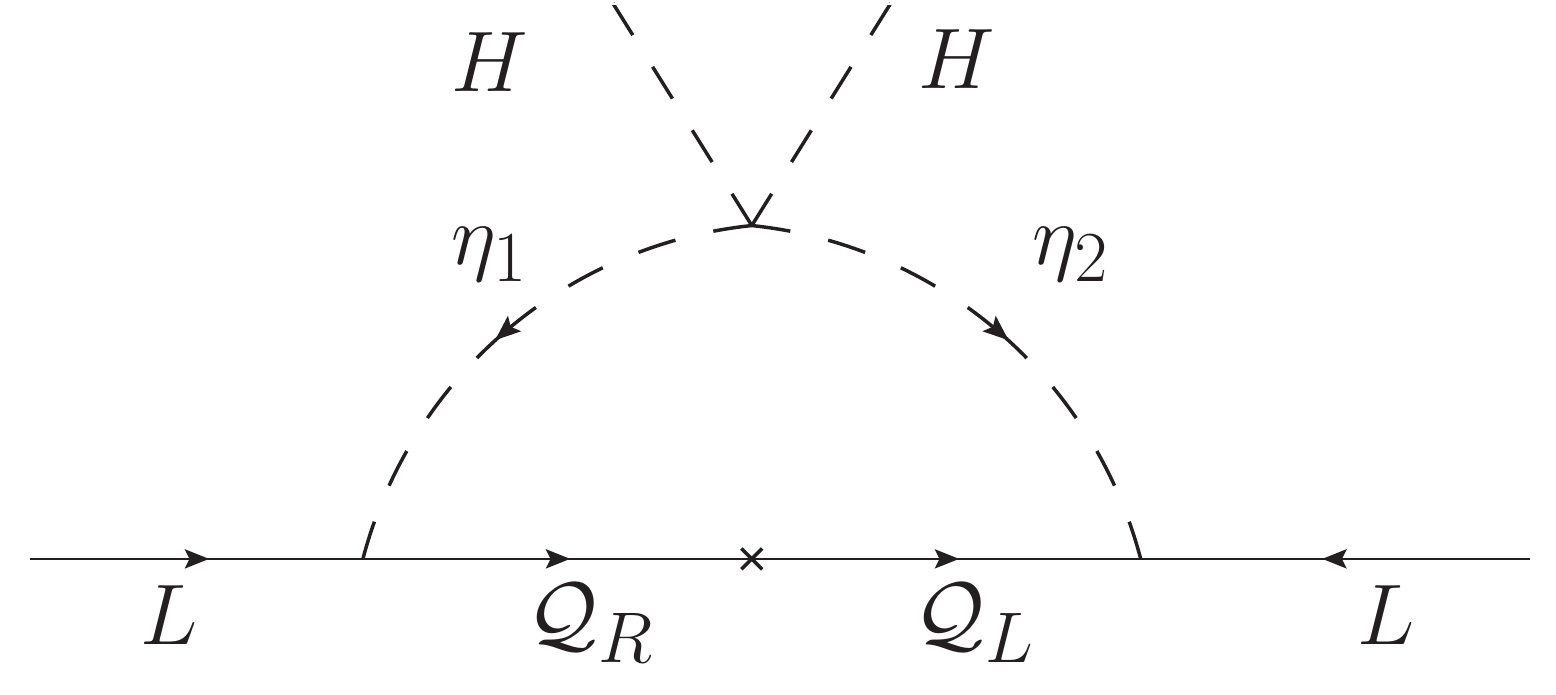}
\caption{$U(1)_D$ flux circulating  the one-loop neutrino mass.}
\label{fig:oneloop}
\end{figure}

The Higgs boson is the same as that in the standard model. The neutral scalar components $\eta^0_a$ of the dark charge carrying scalars $\eta^a$, with cuadratic mass coefficients $\mu_{a}$ in Fig.~\ref{fig:oneloop},  mix into two complex mass eigenstates $S_a$. 
Since the tree-level mass term is forbidden by symmetry, neutrino masses are calculable at one--loop order, by the Feynman diagram displayed in~Fig.~\ref{fig:oneloop}.

One finds the following effective mass matrix 
\begin{align}\label{eq:numassmatrix}
  \left(\mathcal{M}_\nu\right)_{ij}=&N_c\frac{M_{\mathcal{Q}}}{32\pi^2}\left(   h_{i} y_{j} + h_{i} y_{j}  \right)
                                      \frac{\lambda_{\eta_1\eta_2 H} v^2}{m^2_{S^0_2}-m^2_{S^0_1}}
 \left[F\left(\frac{m^2_{S^0_{2}}}{M_{\mathcal{Q}}^2}\right)-F\left(\frac{m^2_{S^0_{1}}}{M_{\mathcal{Q}}^2}\right)\right].
\end{align}
\black
where $F(m_{S_\beta}^2/M_Q^2)=m_{S_\beta}^2\log(m_{S_\beta}^2/M_Q^2)/(m_{S_\beta}^2-M_Q^2)$ and the $\operatorname{SU}(3)_c$ color factor $N_c$ is assumed to be $8$, since the new particles running in the loop transform as octets. Note that the effective one-loop induced neutrino mass matrix has in general rank two, implying two non-vanishing neutrino masses (similar general structure occurs also in non-colored models~\cite{Hehn:2012kz,Ma:2013yga,Hagedorn:2018spx}), as required to account for neutrino oscillation data.

As a simple numerical estimate, assuming $\mu_{\eta_1}^2\gg M_{\mathcal{Q}}^2$, we consider the case $\mu_{\eta_1}^2=\mu_{\eta_2}^2\gg\lambda_{\eta_1\eta_2 H}v^2$ and $\lambda_{3\eta H},\lambda_{4\eta H}\ll1$. Taking $m_{S_2}^2-m_{S_1}^2=\lambda_{\eta_1\eta_2 H}v^2$ and $m_{S_{2R}^a}^2+m_{S_{1R}^a}^2=2\mu_{\eta_1}^2$, we find
\begin{align}
 &\left(\mathcal{M}_\nu\right)_{ij} 
 \sim 0.04\,\mbox{eV}\left(\frac{M_{\mathcal{Q}}}{12.5\, \mbox{TeV}}\right)\left(\frac{ \lambda_{\eta_1\eta_2 H}v^2}{0.1\, \mbox{GeV}^2}\right)\left(\frac{15\,\mbox{TeV}}{\mu_{\eta_1}}\right)^2\left(\frac{h_{i} y_{j}}{10^{-6}}\right).
\end{align}
Small neutrino masses are protected by the small parameter $\lambda_{\eta_1\eta_2 H}$ in whose absence the theory acquires a larger symmetry. As a result, adequate neutrino masses arise naturally for reasonable values of the Yukawa couplings and scalar potential parameters.
Due to the flavor structure of the neutrino mass matrix (Eq. (\ref{eq:numassmatrix})), one can express five of the six Yukawa-couplings $h_i$ and $y_i$ in terms of the neutrino observables \cite{Longas:2015sxk}. Choosing $h_1$ as a free parameter we have 
\begin{align}
&h_i=|h_1|\frac{A_i}{\beta_{11}}, \hspace{1cm}y_i=\frac{1}{2\zeta}\frac{\beta_{ii}}{h_i},
\end{align}
where 
\begin{align*}\label{eq:yukNH}
\beta_{ij} = e^{i\alpha} m_2V_{i2}^*V_{j2}^*+ m_3V_{i3}^*V_{j3}^*,\,\,\,& A_{j} = \pm\sqrt{-e^{i\alpha} m_2m_3(V_{12}^*V_{j3}^*-V_{13}^*V_{j2}^*)^2e^{i2\text{Arg}(h_1)}}+\beta_{1j}e^{i\text{Arg}(h_1)},\,\,\,  \mbox{for NO}, \\
\beta_{ij} = m_1V_{i1}^*V_{j1}^*+ e^{i\alpha} m_2V_{i2}^*V_{j2}^*,\,\,\,& A_{j} = \pm\sqrt{-e^{i\alpha} m_1m_2(V_{11}^*V_{j2}^*-V_{12}^*V_{j1}^*)^2e^{i2\text{Arg}(h_1)}}+\beta_{1j}e^{i\text{Arg}(h_1)},\,\,\, \mbox{for IO}, \\
&\zeta = N_c\frac{M_{\mathcal{Q}}}{32\pi^2}\frac{\lambda^{12}_{\eta_1\eta_2 H} v^2}{m^2_{S^0_2}-m^2_{S^0_1}}
 \left[F\left(\frac{m^2_{S^0_{2}}}{M_{\mathcal{Q}}^2}\right)-F\left(\frac{m^2_{S^0_{1}}}{M_{\mathcal{Q}}^2}\right)\right].
\end{align*}
Note that, in the diagonalization condition $U^{\text{T}}\mathcal{M}_{\nu}U ={\rm diag}(m_1,m_2,m_3)$ we have used the PDG form~\cite{Agashe:2014kda} of the lepton mixing matrix~\cite{Schechter:1980gr}, namely $U=VP$ and $P=\mbox{diag}(1,e^{i\alpha/2},1)$. 

\section{Stability of bound-state dark matter and neutrino masses}
\label{sec:dark-matt-stab}

As suggested in~\cite{Ma:2006km} the radiative nature of neutrino mass generation is used to ensure dark matter stability. Moreover, this idea is combined with the proposal that dark matter emerges in the form of stable neutral hadronic thermal relics, as a neutral bound-state of colored constituents, such as $\mathcal{Q}\mathcal{Q}$, where \Q is a vector-like color octet isosinglet fermion (for a general discussion of the cosmology of a stable colored relic see~\cite{Geller:2018biy}).  
A necessary and sufficient condition for dark matter stability in this case is the presence of a global $U(1)_D$
dark baryon number \footnote{Quantum gravity effects may require to gauge global symmetries \cite{Krauss:1988zc}. An extension where $U(1)_D$ is promoted to a local symmetry can be easily envisaged in a way similar to that considered in~\cite{Agrawal:2016quu}.}, under which the \Q is
charged~\cite{DeLuca:2018mzn}. In our present model construction such symmetry also gives rise to radiative neutrino masses. 
In fact, both dark matter stability, and the Dirac nature of the exotic fermion \Q are equivalent, resulting from dark charge conservation. 

Note that, by construction, the new Yukawa interactions in Eq.~(\ref{eq:lag}) do not affect the stability of our bound state dark matter, since the colored scalars are heavier than $\mathcal{Q}$, and $\mathcal{Q}$$\mathcal{Q}$ annihilation processes mediated by $\eta$ are forbidden by the conserved dark symmetry.
An adequate thermal relic density of bound-state dark matter requires the lightest constituent vector-like color octet Dirac fermion, $\mathcal{Q}$, to have a mass 
$\approx$ 12.5~TeV, so that the $\mathcal{Q}\mathcal{Q}$ hadron weighs approximately 25~TeV~\cite{DeLuca:2018mzn}. 

\begin{figure}[h]
\centering
\includegraphics[scale=0.6]{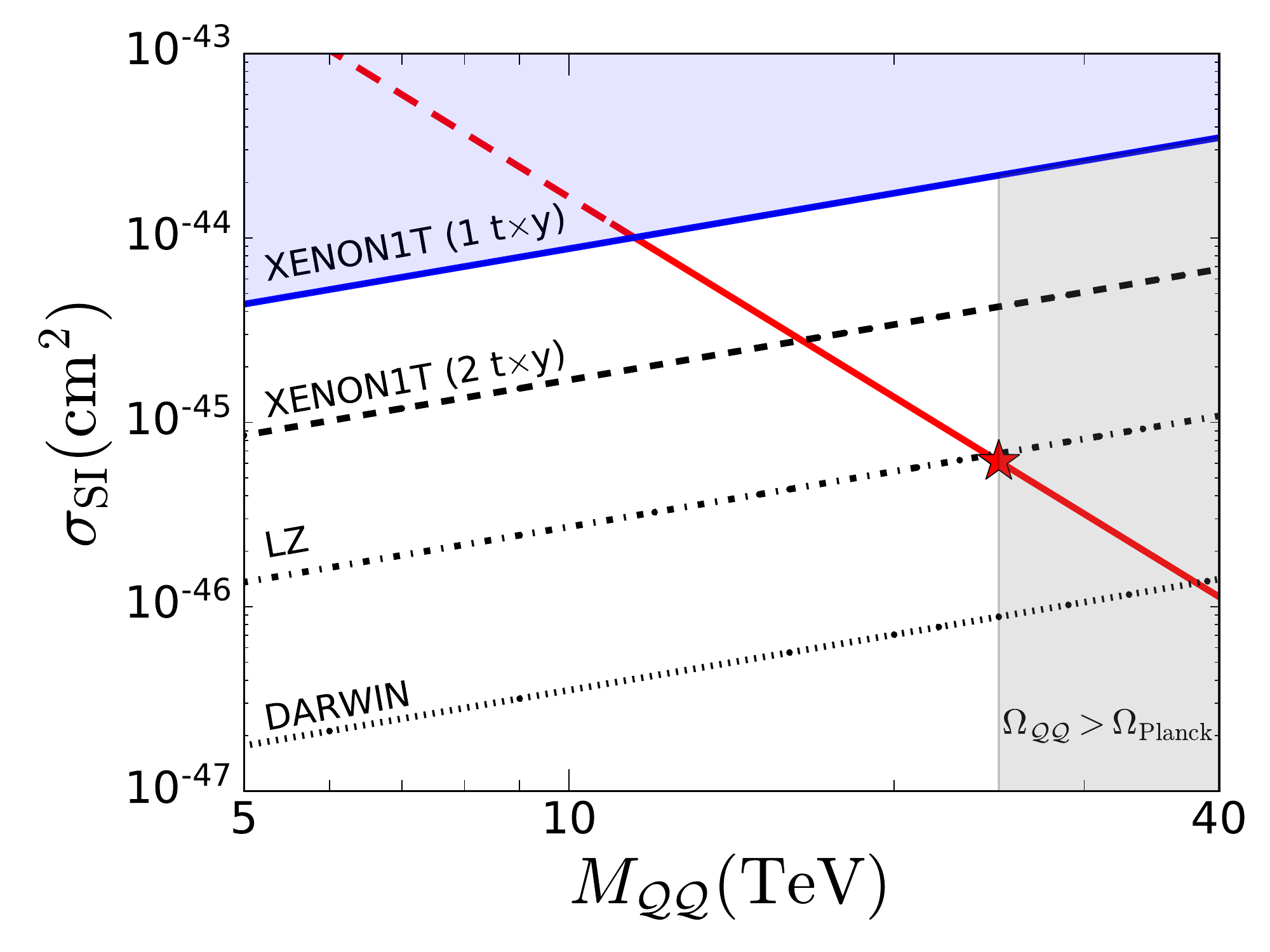}
\caption{Spin-independent cross section as a function of $M_{\mathcal{Q}\mathcal{Q}}=2M_\mathcal{Q}$ (red). The star represents the mass required for a thermal bound state 25~TeVdark matter. Lower values can be probed by direct searches, the current bound is indicated in blue, while the black lines (dashed, dotted and dot-dashed) correspond to future sensitivities.  }
\label{fig:spin-independent}
\end{figure}

Bound-state dark matter made up by our Dirac octets will be seen in direct searches for nuclear recoil at underground dark matter experiments. The relevant spin-independent cross-section is given as
\begin{equation}
\sigma_{\text{SI}}\approx 5.2\times 10^{-46}\ \text{cm}^2\left(\frac{25\ \text{TeV}}{M_{\mathcal{Q}\mathcal{Q} }}\right)^6 \frac{\Omega_{\mathcal{Q}\mathcal{Q}}}{\Omega_{\text{Planck}}}\,,
\end{equation}
and depends rather sensitively on the dark matter mass $M_{\mathcal{Q}\mathcal{Q}}=2M_\mathcal{Q}$, as shown in the red line in Fig.~\ref{fig:spin-independent}.

Note that the star in the figure assumes that the bound-state DM makes up 100\% of the cosmological dark matter.
If an additional dark matter component is present, e.g. made of axions, then bound-state dark matter masses below 25~TeV become allowed, as indicated by the red line.
In this case their spin-independent cross section would be larger, though their share in the relic density will be lower.
The blue line denotes the current Xenon1T limit after 1.0 t$\times$yr exposure~\cite{Aprile:2018dbl}.
This should be compared with the future sensitivities expected at XENON1T~\cite{Aprile:2015uzo}, LZ~\cite{Akerib:2018lyp} and DARWIN~\cite{Aalbers:2016jon}~indicated by the black (dashed, and dot-dashed) lines. \black 
  We also note that, within the standard thermal cosmological scenario, DM masses above 25~TeV are ruled out by current observations of the Planck collaboration~\cite{Ade:2015xua} (gray band).
  Notice also that the current LHC limit of 2~TeV~\cite{CMS:2016ybj,ATLAS:2018yey}, implies that the cross section is always small enough so as to have the bound-state dark matter candidate reaching underground detectors.

  \section{Neutrinoless double beta decay}

In contrast to the proposal in Ref.~\cite{Reig:2018mdk}, here total lepton number is a broken symmetry, hence we expect neutrinoless double beta decay to occur. The effective mass parameter characterizing the amplitude for neutrinoless double beta decay is given by \cite{Rodejohann:2011vc}  
\begin{equation}
\vev{m_{ee}}=\left|\sum_jU_{ej}^2m_j\right|=\left|c_{12}^2c_{13}^2m_1+s_{12}^2c_{13}^2m_2e^{2i\phi_{12}}+s_{13}^2m_3e^{2i\phi_{13}}\right|\,,
\end{equation}
where $m_i$ are the neutrino masses, $c_{12}$ and $s_{13}$ correspond to the angles measured from oscillations and $\phi_{12}$, $\phi_{13}$ are the Majorana phases (here we use the symmetrical parametrization of the lepton mixing matrix~\cite{Schechter:1980gr}).

Since our model predicts the lightest neutrino to be massless, $m_1=0$, it follows that there is one single physical (relative) Majorana phase $\phi\equiv\phi_{12}-\phi_{13}$. Furthermore, one can also write the three physical masses directly in terms of the squared mass splittings measured in neutrino oscillation experiments. Depending on the ordering these masses read
\begin{equation}
\begin{split}
  &{\rm NO}:\,\,m_2=\sqrt{\Delta m^2_{21}}\,,\,\,m_3=\sqrt{\Delta m^2_{31}}\,,\\
  &{\rm IO}:\,\,m_1=\sqrt{\Delta m^2_{13}}\,,\,\,m_2=\sqrt{\Delta m^2_{13}+\Delta m^2_{12}}\,.
\end{split}
\end{equation}

\begin{figure}[!h]
  \includegraphics[scale=0.35]{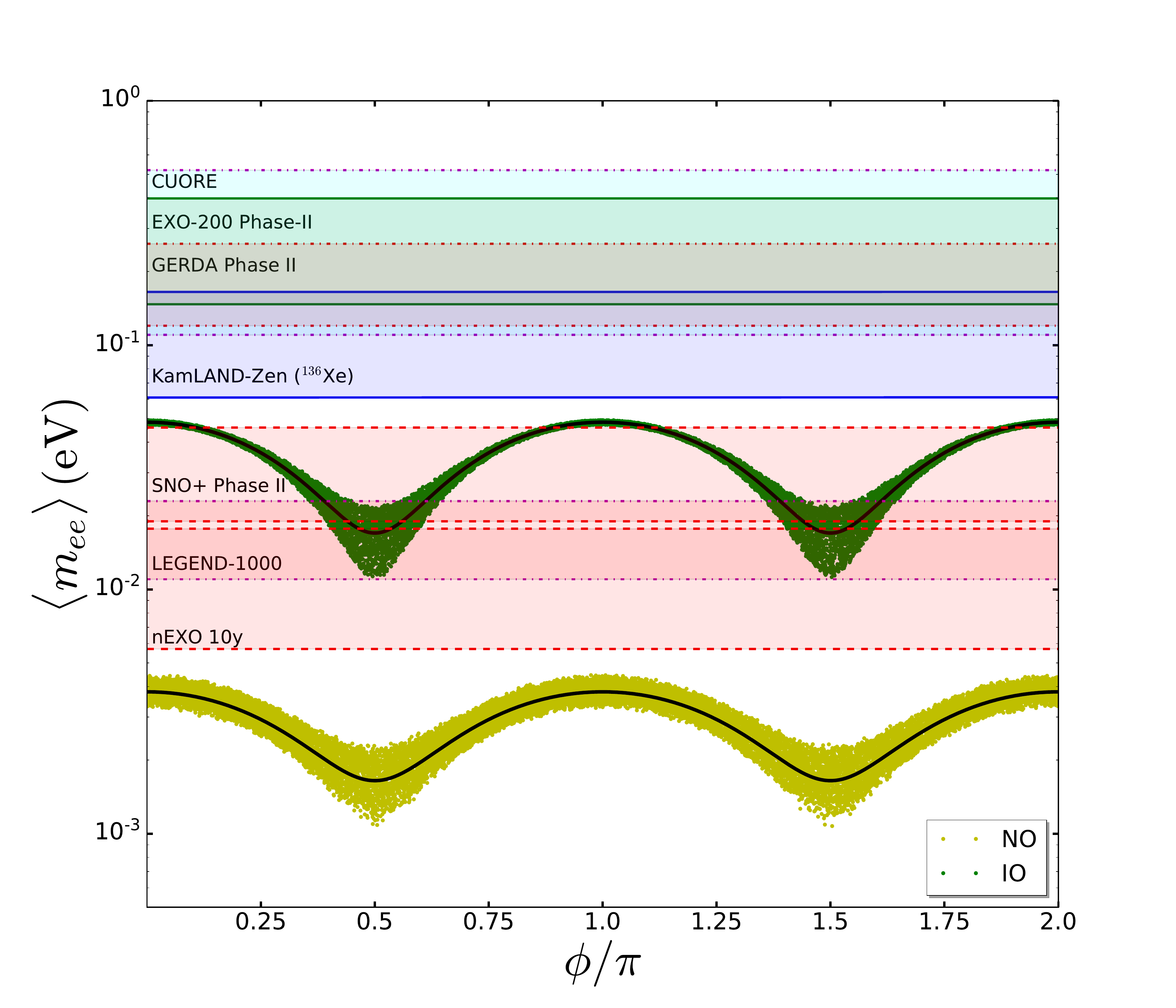}
  \caption{Effective Majorana mass as a function of the Majorana phase for normal (yellow band) and inverted (green band) mass orderings. The bands represent the $3\sigma$ uncertainties in the neutrino oscillation parameters~\cite{deSalas:2017kay,globalfit}. The solid black lines represent their best fit values. The horizontal bands are the experimental limits and sensitivities. }
  \label{fig:mee}
\end{figure}

Given that the lightest neutrino is massless, one can plot the effective Majorana mass parameter $\vev{m_{ee}}$ as a function of the unknown Majorana phase $\phi$, without loss of generality, see Fig.~\ref{fig:mee}. 
The results are given for the cases normal (yellow band) and inverted (green band) mass orderings, varying the neutrino oscillation parameters within $3\sigma$ \cite{deSalas:2017kay,globalfit} of their best fit values. One sees that, in contrast to the general three-neutrino scenario, here the \znbb amplitude never vanishes, even when the neutrino mass ordering is of the normal type (models with this feature typically require the existence of specific flavor symmetries~\cite{Hirsch:2005mc,Dorame:2011eb,Dorame:2012zv}). 
The top four horizontal bands represent the 90\% C.L. upper limits from CUORE \cite{Alduino:2017ehq}, EXO-200 Phase II \cite{Albert:2017owj}, GERDA Phase II \cite{Agostini:2018tnm} and KamLAND-Zen \cite{KamLAND-Zen:2016pfg} experiments.
The sensitivity bands for the upcoming nEXO experiment after 10 years of data taking~\cite{Albert:2017hjq} as well as for the SNO+ Phase II~\cite{Andringa:2015tza} and LEGEND~\cite{Abgrall:2017syy} experiments are indicated by the horizontal red bands. 

\section{Coupling evolution}

\begin{figure}[!h]
	\includegraphics[scale=0.6]{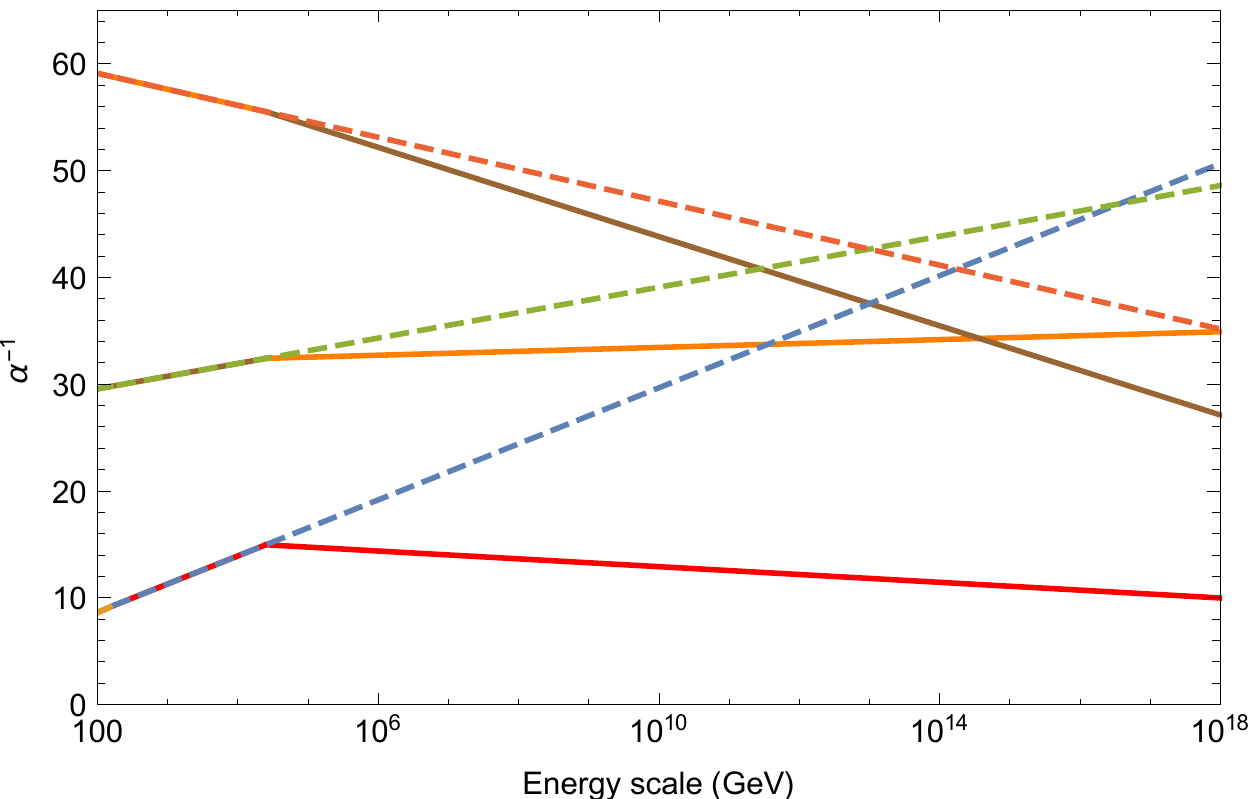}
	\caption{Running of gauge couplings with light scalars: $M_{\eta_a}\approx 25$ TeV. For DM stability, $M_{\eta_a > M_Q}$ is required.}
	\label{fig:LP}
\end{figure}

Given that we have colored multiplets it is interesting to illustrate their effect in the running of coupling constants. At one loop level the evolution of gauge couplings is governed by
\begin{equation}
\alpha_i (\mu_2)^{-1}=\alpha_i (\mu_1)^{-1}-\frac{b_i}{2\pi}\ln \left(\frac{\mu_2}{\mu_1}\right)
\end{equation}
where the $b_i$ coefficients are determined by
\begin{equation}
b_i=-\frac{11}{3}C_2(G)+\frac{2}{3}\sum_fT_{f}d_f+\frac{1}{3}\sum_sT_sd_s      \,.
\end{equation}
For our model, they are given as
\begin{equation}
\begin{split}
&b_3=-7+4\times n_Q+2\times n_\eta\,,\\&
b_2=-\frac{19}{6}+\frac{4}{3}\times n_\eta\,,\\&
b_1=\frac{41}{10}+\frac{4}{5}\times n_\eta\,,
\end{split}
\end{equation}
where $n_\eta$ is the number of colored scalar fields $\eta\sim (8,2,1/2)$. One can easily see that asymptotic freedom can be lost due to the presence of the octets.
However, in our case $n_\eta =2$, hence no Landau pole appears up to the Planck scale, as long as the scalars are heavier than the fermion octets $\mathcal{Q}\sim (\mathbf{8},1,0,1)$. This is illustrated in Fig.~\ref{fig:LP}. 

Notice that the situation displayed in Fig.~\ref{fig:LP} seems to threaten the idea of unification of couplings. This, however, is not the full story. In a Grand Unified Theory (GUT) all the fields come in representations of the GUT group. Complete representations of the GUT group do not spoil unification, but change the value of the coupling constant at $M_{GUT}$. A particular GUT embedding of the color octets $\mathcal{Q}\sim (\mathbf{8},1,0)$, however, lies beyond the scope of this work.

\section{Phenomenology}
\label{sec:color-octets-at}

The cosmological relic abundance requires
$M_{\mathcal{Q}}\approx 12.5$~TeV, which lies beyond the kinematical
reach of the LHC. A future hadron collider with a center of mass
energy of 100~TeV, would probe masses up to
$M_{\mathcal{Q}}\lesssim 15$~TeV~\cite{diCortona:2016fsn,DeLuca:2018mzn}. In
contrast with bound-state dark matter with radiative Dirac neutrino
masses induced by color octets circulating the
loop~\cite{Reig:2018mdk}, in the Majorana case the color octet scalars
can lie at the same scale as the color octet fermion. In fact, if the
lightest color octet scalar, $\eta_1$, transforming as a weak doublet,
has a mass close to that of ${\cal Q}$, they could be pair-produced
with similar cross sections\footnote{We compared the reported cross
  sections of~\cite{diCortona:2016fsn} for the color octet Dirac
  fermion with the MadGraph~\cite{Alwall:2011uj} output of
  FeynRules~\cite{Alloul:2013bka}
  for color-octet isospin-doublet scalar implemented by~\cite{ElHedri:2017nny}.}.
The pair produced scalars further decay into a
$\mathcal{Q}\bar{\mathcal{Q}}$ pair and charged leptons or neutrinos.
This gives rise to similar signals with long-lived bound states but
with extra charged leptons or missing transverse energy.

Note also that our bound-state dark matter constituents can mediate charged lepton flavor violation effects, whose rates will depend on the scalar masses and may reach detectability levels. 

\section{Summary and outlook}
\label{sec:summary-discussion}

We have proposed a simple and viable theory in which dark matter emerges as a stable neutral hadronic thermal relics, whose stability results from an exact $\operatorname{U}(1)_D$ symmetry. Neutrinos pick up radiatively induced Majorana masses from the exchange of colored dark matter constituents, giving a common origin for both dark matter and neutrino mass, with a lower bound for neutrinoless double beta decay and direct tests at direct DM searches at nuclear recoil experiments.
The scheme provides a consistent ultraviolet complete setup, free of Landau poles all the way up to the Planck scale, provided the scalars are heavy
enough. The new states may also lead to other phenomenological signals at future hadron collider as well as \lfv experiments.

\begin{acknowledgments}

Work supported by the Spanish grants FPA2017-85216-P and SEV-2014-0398 (MINECO), by PROMETEOII/2014/084 (Generalitat  Valenciana), by Sostenibilidad-UdeA, and by COLCIENCIAS through the Grants 11156584269 and 111577657253.

\end{acknowledgments}

\bibliographystyle{utphys}
\bibliography{bibliography} 
\end{document}